\documentclass[12pt,preprint]{aastex}

\shorttitle{Optical counterparts of Ultra Luminous X-Ray Sources}
\begin{document}
\title{Optical Counterparts of Ultra Luminous X-Ray Sources}
\author{C. M. Guti\'errez$^1$}
\affil{\small$^1$Instituto de Astrof\'{\i}sica de Canarias, E-38205, La Laguna, 
Tenerife, Spain (cgc@ll.iac.es)}

\newpage
\begin{abstract}
We present optical identification and characterization of  counterparts of four objects
previously catalogued as ultra-luminous X-ray sources. The objects were selected from the Colbert
\& Ptak (2002) catalogue. The optical counterparts are identified as point-like objects with
magnitudes in the range $\sim$17--19.  The optical spectra of three of the sources  (IXO~32, 37
and 40) show the presence of emission lines typical of quasars. The position of these lines allows
a precise estimation of their redshifts (2.769, 0.567  and 0.789 for IXO~32, 37 and 40
respectively). The fourth X-ray source, IXO~35,  is associated with a red object that has a
spectrum typical of an M star in our Galaxy. These identifications  are useful for building clean samples of ULX sources,
selecting suitable targets for  future observations and performing statistical studies on the
different populations of X-ray sources. 

\end{abstract}

\keywords{galaxies: active - quasars: general - X-rays: galaxies}

%% manuscript produces a one-column, double-spaced document:

%% \documentclass[manuscript]{aastex}

%% preprint2 produces a double-column, single-spaced document:

%% \documentclass[preprint2]{aastex}

%% Sometimes a paper's abstract is too long to fit on the
%% title page in preprint2 mode. When that is the case,
%% use the longabstract style option.

\section{Introduction}

One of the most intriguing astrophysical objects are the ultra-luminous X-ray sources (ULXs) that have been
discovered around nearby galaxies by the X-ray satellites {\itshape Einstein}, {\itshape ROSAT}, {\itshape
Chandra} and {\itshape XMM}. Assuming that they are at the same distance as their parent galaxies, their
luminosities in the 0.1--2.4 keV  band are in the range  $10^{39}$--$10^{41}$ erg s$^{-1}$. Several
explanations concerning the nature of these objects in terms of intermediate-mass black holes associated
with globular  clusters, HII regions, supernova remnants, etc.\ (Pakull \& Mirioni 2002; Angelini et al.\
2001; Gao et al.\ 2003; Roberts et al.\ 2003; Wang 2002), local QSOs (Burbidge et al.  2003),  hypothetical
supermassive stars or beamed emission (King et al. 2001; K\"ording et al.\ 2002) have been proposed.
Studies with {\itshape XMM-Newton} (Jenkins et al.\ 2004)  point to a  heterogeneous class of objects whose
spectral properties are similar to those of objects with lower X-ray luminosities. Detailed studies with
{\itshape Chandra} and {\itshape XMM} and the identification of counterparts in other spectral ranges
(optical/infrared/radio) are essential for making progress in the field. The number of such optical 
identifications is still limited (e.g.\ Roberts et al.\ 2001; Wu et al.\ 2002; Liu et al.\ 2004). Some of
the these counterparts have revealed objects at higher redshift than the putative parent galaxies  (Maseti
et al.\ 2003; Clark et al.\ 2005), and are viewed in standard big bang cosmology as contaminated background
objects.

Saveral major compilations of ULX sources exist, those by Colbert \& Ptak (2002) (CP02 hereafter), Swartz et
al. (2004), Liu \& Mirabel (2005) and Liu \& Bregman (2005) (LB05 hereafter). The statistical analysis  of the objects in the
CP02  catalogue made by Irwin et al.\ (2004) indicates considerable contamination by background sources. A
direct confirmation of these and a better understanding on the nature of ULX sources can be achieved by
identifying counterparts in other bands (in particular in the optical, as is widely recognized: see for
instance the review by van der Marel 2004). This motivated us to start this study searching for such
possible optical counterparts in the major existing optical surveys. For example, we have identified
possible optical counterparts in the DSS plates for about $\sim$50\% of the objects compiled by CP02. The
typical magnitudes of such objects are 17--20  in the  $b$  band and are therefore bright enough targets 
for spectroscopic observations with 2 to 4 m telescopes. In previous work (Arp et al.\ 2004: Guti\'errez \&
L\'opez-Corredoira 2005) we have demonstrated the feasibility of such studies and present our first results
with the identification and characterization of nine  such {\itshape ROSAT} ULX sources. In eight cases the
sources look point-like and turned out to be  quasars at high redshift. The remaining object was in a
spiral arm of NGC~1073 and is apparently embedded in an HII region. Here, we present further results with
an analysis of the counterparts of four additional ULX sources.

\section{Sample selection and observations}

\subsection{Imaging}
In this paper we consider the cases of IXO~32, 35, 37 and 40 (we follow the 
notation by CP02). Table~1 summarizes the main properties of such sources (taken from the
compilation by CP02). We look for possible optical counterparts
of these X-ray sources in the Digital Sky Survey (DSS) plates, the USNO
catalogue, and the released Sloan Digital Sky Survey (SDSS) data.  
For the four cases we checked that there are point-like objects compatible with the
X-ray positions. The last column in Table~1 lists the offset between the optical and
X-ray coordinates. The fields around IXO~32, 35 and 37 were also observed with the IAC80
telescope\footnote{The IAC~80 is located at the Spanish Teide Observatory on the
island of Tenerife and is operated by the Instituto de Astrof\'\i sica de
Canarias} in May 2004 and December 2005. For these objects we took single exposures of 1800 s in $BR$ for
IXO~32, 600 s in $BVRI$ for
IXO35, and 600 s in $R$ for IXO~37. The
observations were reduced using IRAF\footnote{IRAF is the Image Reduction and
Analysis Facility, written and supported by the IRAF programming group at the
national Optical Astronomy Observatories (NOAO) in Tucson, Arizona.} following a
standard procedure  The nights were photometric  and
we use Landolt stars (Landolt 1992) to perform an absolute calibration with an
uncertainty of (2$\sigma$) $\leq 0.05$ mag in each
filter. Figure~1 show the images with the identification of the
optical  counterpart. The images (2$'$ $\times$ 2$'$)  are centred on the nominal X-ray positions.
In one case (IXO~32) there is another source slightly shifted (12 arcsec) from
the X-ray coordinates.

\subsection{Spectroscopy}
The spectroscopic observations presented here were taken in February 2004 with
the  WHT\footnote{The William Herschel Telescope (WHT) is operated by the Isaac
Newton Group and the IAC in Spain's Roque de los Muchachos Observatory}. We used
the blue and the red arms  of the ISIS spectrograph with the R300B and
R158R grisms. The slit width was 2 arcsec. We  took a Cu--Ar and Cu--Ne lamp with a
slitwidth of 1 arcsec at the beginning of the night for wavelength calibration.
The stability of the wavelength calibration during the night and pointing was 
checked using the main sky lines. The sampling was 1.71 \AA~and 3.26 \AA~in the
blue and red arms respectively. For each target, a single image was taken with  exposure
times of 1800 s for the counterparts of IXO~32, 37 and 40, and 900 s for the
counterpart of IXO~35. The spectra were analysed following a standard
procedure using IRAF of bias subtraction, extraction of the spectra
and wavelength calibration. We used the standard spectroscopic star  Feige~34
(Oke 1990) to correct for the response of the configuration to different
wavelengths. The star was observed only three times during two  nights and the
slit for the targets was not  positioned at parallactic angles, so this correction
is only indicative.   Given the prohibitive time needed to secure flat field images
(specially in the blue part of the spectra), we did not correct for that effect.
However, we have checked that this correction would be very small ($\leq 1$ \%).
None of these uncertainties is relevant for the analysis and results presented in
this paper.

\section{Analysis}
The spectra of the four objects are presented in Figure~2. The four spectra show  features that allow  clear identification
and characterization. Table~2 lists the main properties of these spectra. The
analysis of each object is presented below.

\subsection{IXO~32}
This source is also listed in the LB05 catalogue as X06 (around the galaxy NGC~2775). The optical images
show two point-like objects with a separation  $\sim$7 arcsec and
distant $\sim$5 and $\sim$12 arcsec respectively from the nominal X-ray position.  Only the
brightest (the object at $\sim$12 arcsec)  is listed in the USNO catalogue. We put the slit
crossing both objects. The spectrum of the bright object has typical absorption lines of a
star being the most prominent absorption features, the H \& K CaII, and the Balmer lines.
The fainter  is very blue and turn out to be  an active galactic nuclei/quasar. Figure~2 shows the spectrum of
this object in which  the broad emission Ly-$\beta$+OVI, Ly-$\alpha$, SiIV+OIV (1400 \AA),
CII (1549 \AA) and  CIII (1909 \AA) lines are obvious. From the position of the line CII 
(1549 \AA) we estimate a redshift $z=2.769$.  After the spectroscopic observations we
discovered that the field has been observed with SDSS and that the object situated at 3.2
arcsecs from the nominal CP02 X-ray position has been catalogued as a star with with
magnitudes $r=18.62$ mag and $g=18.86$ mag. 

\subsection{IXO~35}
This object is also listed in the LB05 catalogue and denoted as NGC~3226 X03.
The USNO catalogue lists an object with magnitudes  $b=18.8$ mag and $r=17.2$ mag  at 2.8
and 2.1 arcsec from the CP02 and LB05 X-ray positions respectively.  The
observations taken with the IAC80 telescope show a point-like object with
magnitudes  18.87, 17.97, 16.65 and 14.79 in the $B$, $V$, $R$ and $I$ bands respectively.
The nearest neighbour listed in the USNO catalogue and detected with the IAC80 at
$R=19.3$ mag is $\sim$27 arcsec SE.  The source is also listed in 2MASS
with magnitudes 13.24, 12.70 and 12.39 in the $J$, $H$ and $K$ bands respectively.
According to the maps by Schlegel et al.\ (1998), possible corrections for galactic
extinction are below 0.1 mag in the blue band.

The optical spectrum of this object is dominated by strong absorption bands (VO,
Na and TiO) typical of a cold star. We detect also the Ca H and K and  Balmer 
emission lines ($H\alpha,~H\beta,~H\gamma,~H\delta$ and $H\epsilon$), although some
of them are in the middle of absorption bands, making the estimation of
equivalent widths uncertain. Following the calibration by Hawley et al.\
(2002), based on several photometric and spectroscopic indices, we classify the
object as an M3--6 star. From this and with the magnitude in the $J$ band it is
possible to estimate the distance and then the X-ray luminosity. The resulting
distance is in the range 41 pc (for an M6 type) to 157 pc (for an M3 type), which
corresponds to X-ray luminosities of $8.6\,10^{27}$ and $1.3\,10^{29}$ erg
s$^{-1}$ respectively. These luminosities are within the range found by Schmitt \& Liefke
(2004) for a volume-limited sample of nearby M stars.

\subsection{IXO~37}
The object is also listed in the LB05 catalogue with the name X02 around the galaxy IC~2597.
A possible counterpart at a distance of  $\sim$4  arcsec from the X-ray source appears in the
optical images having a magnitude of 19.44 in $R$ in the observations at the IAC80 telescope.
The spectrum is shown in Fig.~2. We
identify the forbidden  narrow emission line  OII($\lambda\lambda$ 3727 \AA) and 
OIII($\lambda\lambda$ 4959, 5007\AA) as the most important features. The redshift is
$z=0.567$. At this redshift the H$\beta$  lies at $\sim$7617 \AA, which we identify as a bump
in the middle of a telluric  line. We also identify $H\alpha$  at 10294 \AA~in a spectral
region (not shown in the figure)  of low sensitivity of the detector and severely
contaminated by sky emission lines. Other emission lines in the red arm are from NeIII
($\lambda 3869$). Correcting the spectrum for redshift, we detect in the blue arm the
MgII($\lambda 2799$ \AA) emission line. The impossibility to measure the flux of the H$\beta$
and H$\alpha$ lines precludes the application of the common diagnostic  diagrams. However,
based on the X-ray emission, we classsify this object as a Seyfert I, AGN

\subsection{IXO~40}
DSS plates show only a possible optical counterpart of this ULX at $\sim$2
arcsec from the nominal X-ray position  and with magnitudes 17.9 and 19.1 in $r$ and $b$ respectively. 
The optical spectrum shows  emission lines typical of AGN/QSOs. The
main features are   broad emissions  of CIII(1909 \AA) and MgII (2799 \AA) in the
blue arm. OII($\lambda\lambda$ 3727 \AA), NeIII($\lambda\lambda$ 3869 \AA) and possibly OIII($\lambda\lambda$
4959, 5007\AA) are detected in the red. The resulting redshift is $z=0.789$. 

\section{Discussion and conclusions}

The poor spatial resolution of {\itshape ROSAT} images is irrelevant for the optical
identifications presented here. In fact, as discussed in Guti\'errez \&
L\'opez-Corredoira (2005), the low density of bright quasars ($\sim$2-3 per square
degree brighter than 19 mag) makes unlikely a chance projection between the X-ray
sources and the quasars identified as the optical counterparts of IXO~32, 37 and 40. A
similar argument can be applied in the case of IXO~35: from the local density  of M
stars (Martini \& Osmer 1998), we have estimated that the probability to have randomly
a source at $\sim 3$ arcsecs from the X-ray nominal position is below $\sim 10^{-6}$.
These simple arguments confirm the reliability of our identifications. The
sample analyzed in this paper suffer of several biases and then can not be 
considered as statistical representative of the whole sample of ULXs listed in the
CP02 catalogue. In fact, the objects selected  are restricted to those with a bright
point-like optical counterpart ($\sim 19$ mag). So, for instance we have excluded a
priori the possibility to detect X-ray binary stars within the parent galaxy. The
objects were also selected to be in relatively isolated regions that allow an
unambiguous identification. This is against the detection of objects within star
forming complexes. The statistical implications of these and other
identifications are in progress and will be addressed in a forthcoming paper
(L\'opez-Corredoira \& Guti\'errez A\&A submitted). In any case the results presented
here reinforce the importance of multiwavelength studies of these X-ray sources as one
of the most promising ways to disentangle the nature of these objects and for the
construction of clean samples of ULX sources for further studies.

\newpage

\section*{Acknowledgements}
The author is especially grateful to M. L\'opez-Corredoira, a close collaborator
in this project, for useful suggestions and
comments. We thank also J. A. Caballero for useful hints about the properties of
the M star identified as possible counterpart of IXO~35. The author was supported  by the 
{\it Ram\'on y
Cajal} Programme of the Spanish science ministery.

\clearpage

\begin{deluxetable}{lccccccccc}
\tablewidth{0pt}
\tabletypesize{\scriptsize}
\tablecaption{Optical and X-ray properties of ULXs}
\tablehead{
\colhead{ID} & \colhead{RA (J2000)}  & \colhead{Dec (J2000)}  & \colhead{$\log\,(L_X)$}   
& \colhead{ID Gal.}  & \colhead{Type}  & \colhead{$D$} &
\colhead{$d$} & \colhead{$d/R_{25}$} & \colhead{$\Delta$}\\
 & \colhead{(hh:mm:ss.s)} 
 & \colhead{($^\circ$:$^\prime$:$^{\prime \prime}$)} & \colhead{(erg
 s$^{-1}$)} &  & & \colhead{(Mpc)} & \colhead{(arcmin)} &  & \colhead{(arcsec)}\\
}
\startdata
IXO~32 & 09 10 19.9 &	+07 06 00 &  39.2 & NGC 2775	& +2.0	        & 18.05	& 3.7 & 1.7 & 5\\
IXO~35 & 10 23 26.0 &	+19 56 20 &  39.2 & NGC 3226	& $-$5.0	& 17.63	& 2.4 & 1.6 & 3\\
IXO~37 & 10 37 39.6 &	$-$27 05 23 &  40.9 & IC 2597	& $-$4.0	& 40.09	& 1.8 & 1.4 & 4\\
IXO~40 & 11 50 57.9 &	$-$28 44 02 &  39.9 & NGC 3923	& $-$5.0	& 22.24	& 4.4 & 1.2 & 2\\
\enddata
\tablenotetext{*}{
1. Identification of the ULXs; 
2-3. RA (J2000) and Dec (2000) positions; 
4. Log of luminosity in the band 0.1-2.4 KeV  assuming 
that the X-ray sources are at the distance of the parent galaxy; 
5-7; Identification, morphological type and distance of the parent galaxy; 
8-9. Angular distance (in arcmin and in units of $R_{25}$ of the central galaxy) between 
these galaxies and the X-ray source; 10. Difference
in arcsecs between X-ray and optical nominal coordinates. Columns 1-9 have been taken from Colbert \&
Ptak (2002).}
\end{deluxetable}

\clearpage

\begin{deluxetable}{lcccc}
\tablewidth{0pt}
\tabletypesize{\scriptsize}
\tablecaption{Properties of optical counterparts}
\tablehead{
\colhead{ID} & \colhead{Magnitudes}  & \colhead{Main Features}   
& \colhead{Counterpart} &\colhead{Redshift} \\
}
\startdata
IXO~32 & 18.47 & Ly-$\beta$+OVI, Ly-$\alpha$, CII, SiIV+OIV, CIII             & QSO    & 2.769          \\
IXO~35 & 16.65 & Molecular bands       & M star & 41-157 pc     \\
IXO~37 & 19.44  & OII, OII, MgII        & AGN    & 0.567         \\
IXO~40 & 17.9  & CIII, MgII, OII, NeIII,OIII & QSO    & 0.789         \\
\enddata
\tablenotetext{*}{
1. Identification of the ULXs following the Colbert \& Ptak (2002)
notation; 2. Magnitude of optical counterparts in the $R$ band (except for IXO~40 which are 
photographic magnitude in
the red plates of DSS); 3. Main spectral features; 4. Type: 
5. Redshift (or distance for the case of IXO~35).}
\end{deluxetable}

\clearpage

\begin{figure}
\begin{center}
\includegraphics[angle=0,scale=.90]{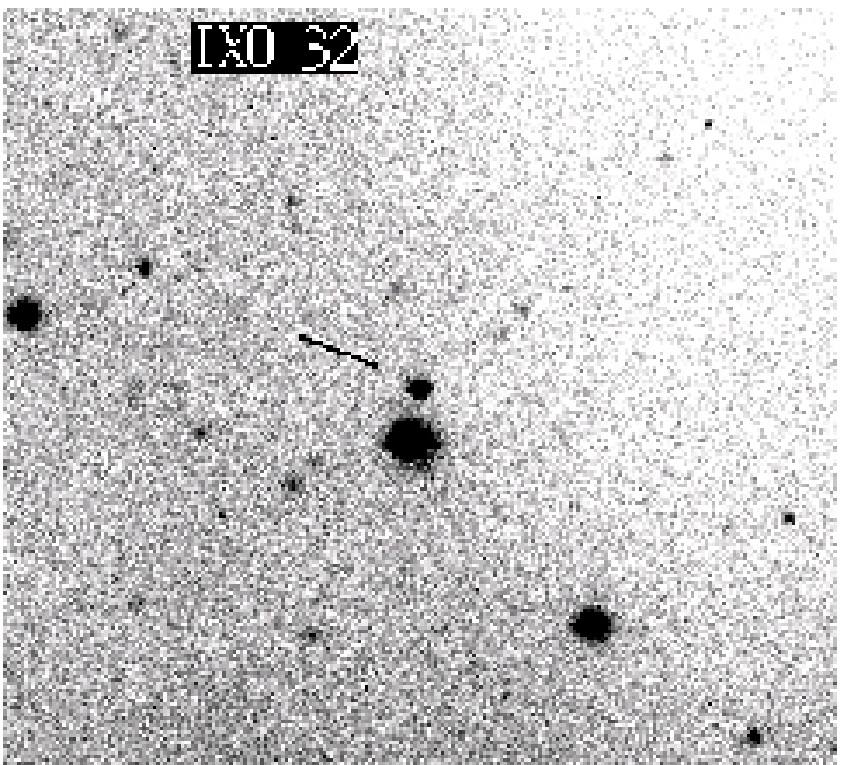}
\includegraphics[angle=0,scale=.60]{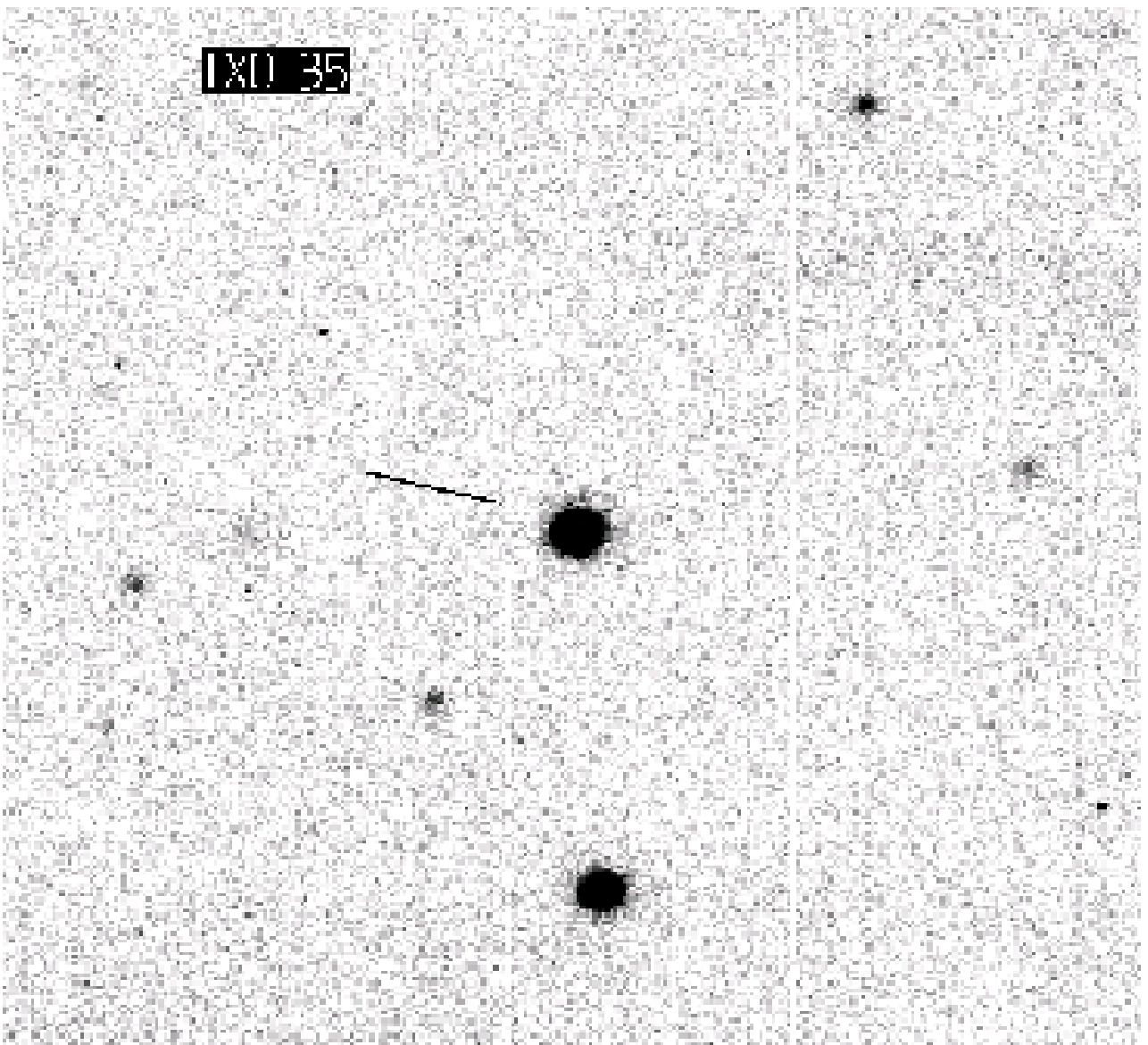}
\includegraphics[angle=0,scale=.90]{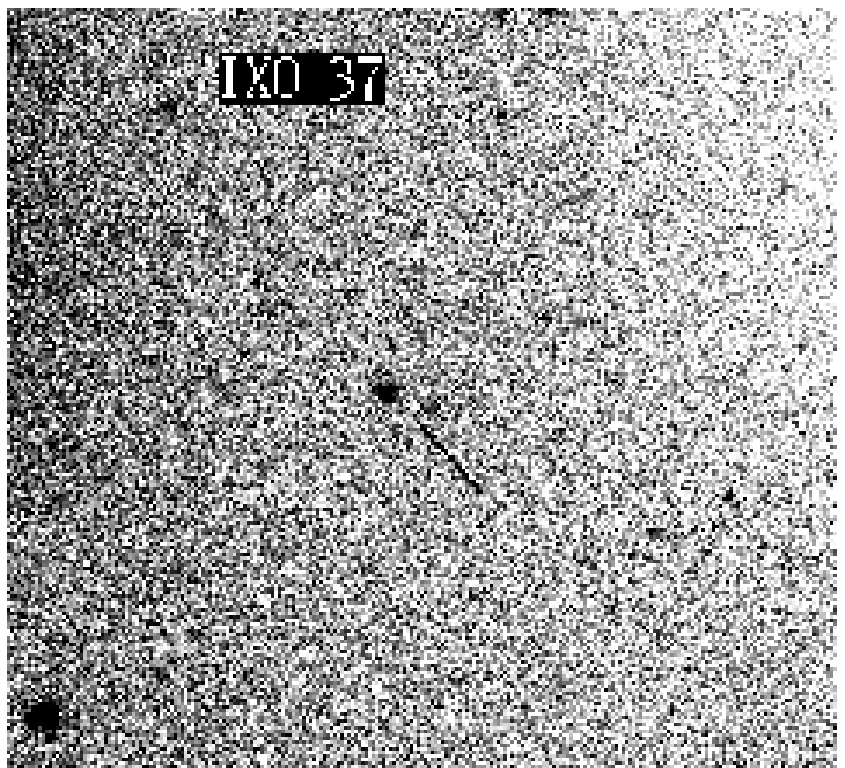}
\includegraphics[angle=0,scale=.66]{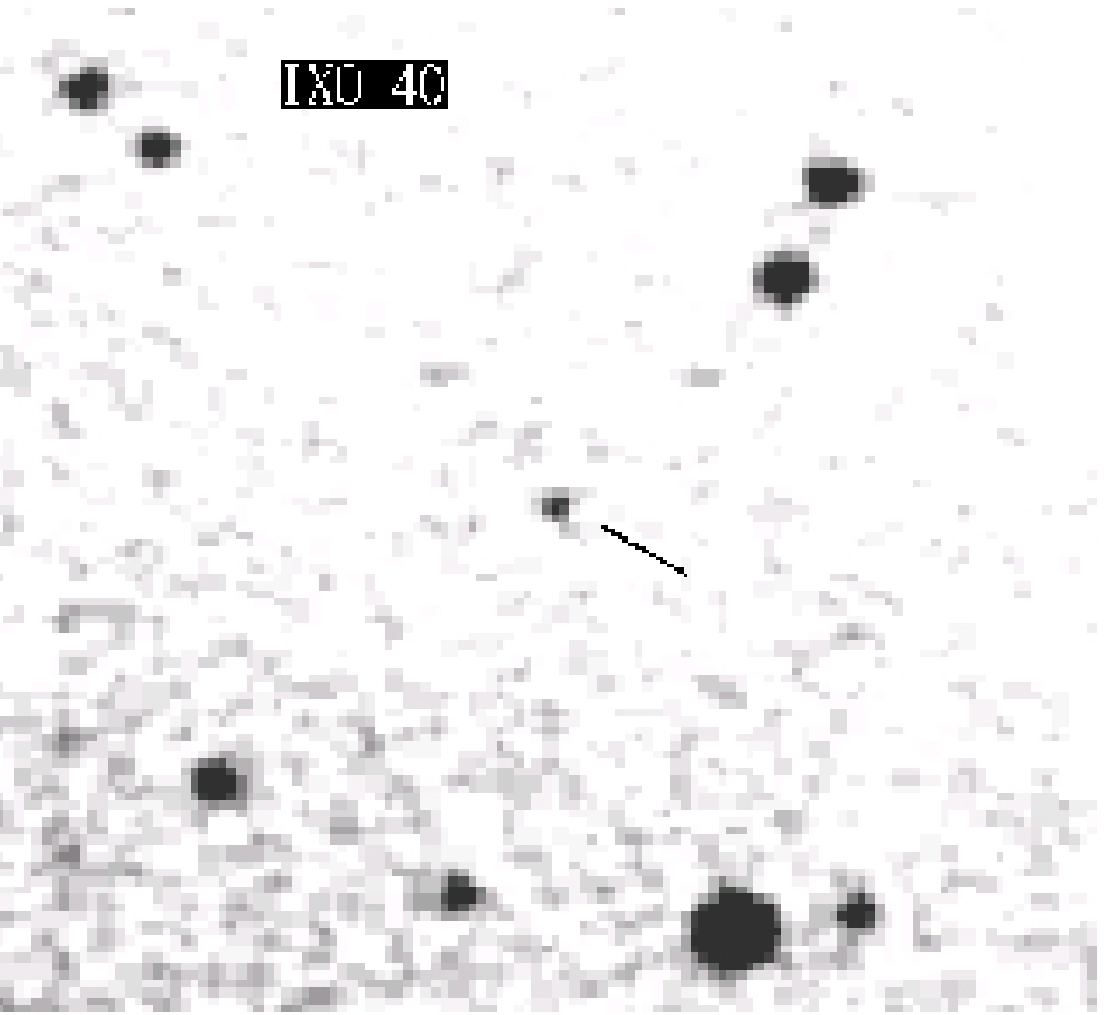}
\end{center}
\caption{Optical images of 2 square arcminutes centred on X-ray positions of the sources 
analysed in this paper. For object IXO~40 the image corresponds to the
red filter of the DSS plates; the other images were obtained with the IAC80
telescope in the $R$  (IXO~32 and 37), and $I$ (IXO~35) filters respectively. The small lines  identify the optical counterpart of the X-ray
sources.  Names according to the notation by Colbert \& Ptak (2002) are indicated. North is up
and east to the left.}
\end{figure}

\clearpage

\begin{figure} 
\includegraphics[angle=0,scale=.82]{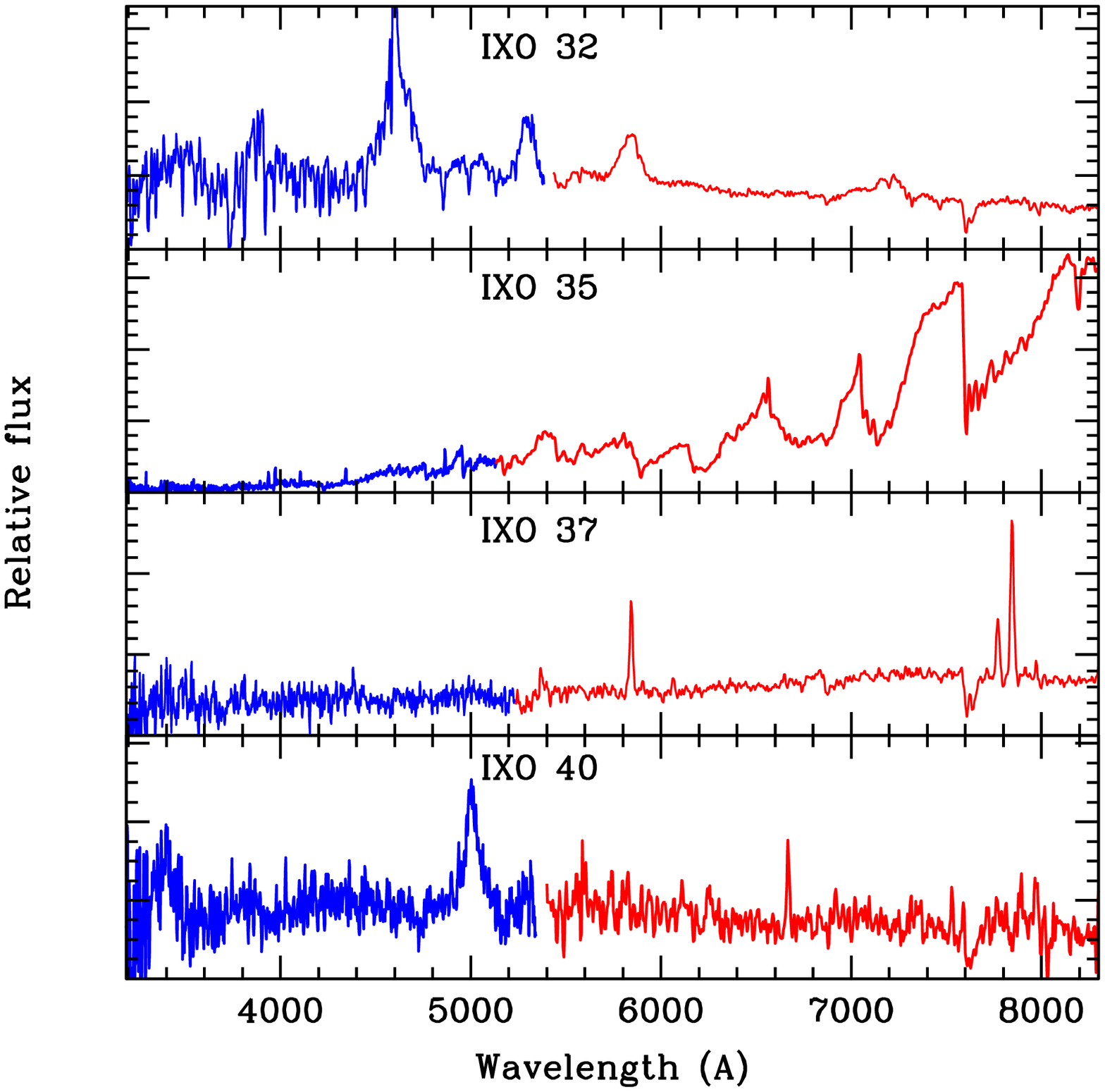}
\caption{Optical spectra  of the counterpart of ULX sources analysed in this paper.
 The $y$-axis is the flux in arbitrary units. The absoption features centered at $\sim 6875$ and $\sim
 7610$ \AA~are telluric bands due to molecular oxigen.}  
\end{figure}


\clearpage

\begin{thebibliography}{}

\bibitem []{} Angelini, L., Loewenstein, M.
\& Mushotzky, R. F.  2001, ApJ, 557, L35

\bibitem []{}  Arp, H., Guti\'errez, C. M. \&
L\'opez-Corredoira, M. 2004, A\&A , 418, 877

\bibitem []{} Burbidge, E. M., Burbidge, G. R. \& Arp H. C. 2003,  
A\&A, 400, L17

\bibitem []{} Clark, D. M. et al. 2005 ApJ, 631, L109

\bibitem []{} Colbert, E. \& Ptak, A. 2002, A
 ApJS, 143, 25

\bibitem []{} Gao, Y., Wang, Q. D., Appleton, P. N. \& Lucas, R. A. 2003,  ApJ, 596, L171 

\bibitem []{} Guti\'errez, C. M.\& L\'opez-Corredoira. M. 2005, ApJ, 622, L89

\bibitem []{} Hawley, S. L. et al. 2002, AJ, 123, 3409

\bibitem []{} Irwin, J. A., Gregman, J. N. \& Athey, A. E. 2004, ApJ, 601, L143

\bibitem []{} Jenkins, I. P., Roberts, T. P., Warwick, R. S., Kilgard, R. E. \&
Ward, M. J. 2004, MNRAS, 349, 404

\bibitem []{} King, A. R., Davies, M. B., Ward, M.
J., Fabbiano, G. \& Elvis, M. 2001,  ApJ,
552, L109
 
\bibitem []{} K\"ording, E., Falcke, H. \& Markoff, S. 2002, A\&A, 
382, L13

\bibitem []{} Landolt, A. U. 1992, A\&A, 104, 340

\bibitem []{} Liu, J. F. \& Bregman, J. N. 2005

\bibitem []{} Liu, J. F., Bregman, J. N. \& Seitzer, P. 2004, ApJ, 602, 249

\bibitem []{} Liu, Q.-Z., \& Mirabel, I. F. 2005, A\&A, 429, 1125

\bibitem []{} L\'opez-Corredoira, M. \& Guti\'errez, C. M. 2004, A\&A, 421, 407

\bibitem []{} Martini, P., \& Osmer, P. S. 1998, AJ, 116, 2513

\bibitem []{} Maseti, N., Foschini, L., Ho, L. C., Dadina, M., Di Cocco, G., Malaguti,
G. \& Palazzi, E. 2003, A\&A, 406, L27

\bibitem []{}  Oke, J. B. 1990, AJ, 99, 1621

%\bibitem []{} Oke, J. B. \& Gunn, J. E. 1983, ApJ, 266, 713

\bibitem []{} Pakull, M. W. \& Mirioni, L. 2002 (astro-ph/0202488). New Visions of the X-ray Universe 
in the
XMM-Newton and Chandra Era', 26-30 November 2001, ESTEC, The Netherlands
 
\bibitem []{} Roberts, T. P. et al. 2001, MNRAS, 325, L7
 
\bibitem []{} Roberts, T. P., Goad, M. R. \& Warwick, R. S. 2003,  
MNRAS, 342, 709

\bibitem []{} Schlegel, D. J., Finkbeiner, D. P. \& Davis, M. 1998, ApJ, 500, 525

\bibitem []{} Schmitt, J. H. M. M., \&  Liefke, C. 2004, A\&A, 417, 651

\bibitem []{} Swartz, D. A.,  Ghosh, K. K., Tennant, A. F. \& Wu, K. 2004, ApJSS, 154, 519

\bibitem []{} van der Marel, R. P.  2004, in Coevolution of Black Holes and Galaxies, 
ed. L. C. Ho (Cambridge: Cambridge Univ. Press), 37

\bibitem []{} Wang, Q. D. 2002,  MNRAS, 332, 764

\bibitem []{} Wu, H. et al. 2002, ApJ, 577, 738

\end{thebibliography}
\end{document}